%% file: main.tex
\newcommand{\DoPrePrint}{1} 

\ifnum\DoPrePrint=1
  \documentclass[aps,prd,preprint,showpacs,superscriptaddress,nofootinbib,linenumbers]{revtex4}

\else
  \documentclass[aps,prd,twocolumn,showpacs,superscriptaddress,nofootinbib]{revtex4}
\fi

\usepackage{graphicx}% Include figure files
\usepackage{dcolumn}% Align table columns on decimal point
\usepackage{bm}% bold math
\usepackage{units}
\usepackage{multirow}
\usepackage{xspace}
\usepackage[usenames]{color}
\usepackage{amssymb}

\newcommand{\sizecheck}{0} % 0 to do nothing; 1 to check size
\newcommand{\numu}{\ensuremath{\nu_\mu}\xspace}
\newcommand{\numubar}{\ensuremath{\bar{\nu}_\mu}\xspace}

\newcommand{\xf}{\ensuremath{x_{F}}\xspace}
\newcommand{\pt}{\ensuremath{p_{T}}\xspace}
\newcommand{\pz}{\ensuremath{p_{z}}\xspace}
\newcommand{\minerva}{MINERvA\xspace}

\newcommand{\gev}[1]{\unit[#1]{GeV}}
\newcommand{\gevc}[1]{\unit[#1]{GeV/c}}
\newcommand{\pip}{\ensuremath{\pi^{+}}\xspace}
\newcommand{\lownu}{low-\ensuremath{\nu}\xspace}
\newcommand{\invxs}{\ensuremath{E_{\pi} \frac{\mathrm{d}^3\sigma}{\mathrm{d}p^3}}\xspace}
\newcommand{\sigabs}{\ensuremath{\sigma_\mathrm{abs}}\xspace}

\newcommand{\siginel}{\ensuremath{\sigma_\mathrm{inel}}\xspace}

\begin{document}
\ifnum\DoPrePrint=1
%\linenumbers
\fi

\preprint{FNAL-PUB-16-091-ND}

%\title{Neutrino Flux Predictions for the Low Energy NuMI Beam}% Force line breaks with \\
%\author{L.~Aliaga}                        \affiliation{\WM}
%\author{M.~Kordosky}                      \affiliation{\WM}
%\author{T.~Golan}                         \affiliation{\Rochester}  \affiliation{\FNAL}
%
%\collaboration{MINERvA Collaboration}%\noaffiliation

\input{title_and_authors}

\date{\today}% It is always \today, today,

\begin{abstract}
Knowledge of the neutrino flux produced by the Neutrinos at the Main Injector (NuMI) beamline is essential to the neutrino oscillation and neutrino interaction measurements of the MINERvA, MINOS+, NOvA and MicroBooNE experiments at Fermi National Accelerator Laboratory.  We have produced a flux prediction which uses all available and relevant hadron production data, incorporating measurements of particle production off of thin targets as well as measurements of particle yields from a spare NuMI target exposed to a \gev{120} proton beam.  The result is the most precise flux prediction achieved for a neutrino beam in the one to tens of GeV energy region.  We have also compared the prediction to {\em in situ} measurements of the neutrino flux and find good agreement. 
\end{abstract}


\ifnum\sizecheck=0  
\maketitle
\fi

%\tableofcontents

%\section*{Introduction}

\input{intro}

\input{hadroproduction}

\input{focusing}

\input{prediction}

\input{insitu}

\input{conclusions}

\ifnum\sizecheck=0
  \input{acknowledgments}
\fi

\bigskip

\ifnum\sizecheck=0
\bibliography{thebib}
\fi

%\clearpage
%%\clearpage
%\input{appendix-oneLoop.tex}
%%% moved into main text (results.tex), ksm 14 Dec 2015
%%%\input{appendix-result.tex}

%\ifnum\PRLsupp=0
%  \clearpage
%\input{appendix.tex}
%\fi

\end{document}

%% file: title_and_authors.tex
%\topmargin=0mm %% No idea WHY I need this.. but I do.

\title{Neutrino Flux Predictions for the NuMI Beam}

%% List of institution addresses, in command form.
%% List of institution addresses, in command form.
\newcommand{\Rutgers}{Rutgers, The State University of New Jersey, Piscataway, New Jersey 08854, USA}
\newcommand{\Hampton}{Hampton University, Dept. of Physics, Hampton, VA 23668, USA}
\newcommand{\Dortmund}{Institute of Physics, Dortmund University, 44221, Germany }
\newcommand{\Otterbein}{Department of Physics, Otterbein University, 1 South Grove Street, Westerville, OH, 43081 USA}
\newcommand{\JMU}{James Madison University, Harrisonburg, Virginia 22807, USA}
\newcommand{\Florida}{University of Florida, Department of Physics, Gainesville, FL 32611}
\newcommand{\UCIrvine}{Department of Physics and Astronomy, University of California, Irvine, Irvine, California 92697-4575, USA}
\newcommand{\CBPF}{Centro Brasileiro de Pesquisas F\'{i}sicas, Rua Dr. Xavier Sigaud 150, Urca, Rio de Janeiro, Rio de Janeiro, 22290-180, Brazil}
\newcommand{\PUCP}{Secci\'{o}n F\'{i}sica, Departamento de Ciencias, Pontificia Universidad Cat\'{o}lica del Per\'{u}, Apartado 1761, Lima, Per\'{u}}
\newcommand{\INRM}{Institute for Nuclear Research of the Russian Academy of Sciences, 117312 Moscow, Russia}
\newcommand{\Jlab}{Jefferson Lab, 12000 Jefferson Avenue, Newport News, VA 23606, USA}
\newcommand{\Pittsburgh}{Department of Physics and Astronomy, University of Pittsburgh, Pittsburgh, Pennsylvania 15260, USA}
\newcommand{\Guanajuato}{Campus Le\'{o}n y Campus Guanajuato, Universidad de Guanajuato, Lascurain de Retana No. 5, Colonia Centro, Guanajuato 36000, Guanajuato M\'{e}xico.}
\newcommand{\Athens}{Department of Physics, University of Athens, GR-15771 Athens, Greece}
\newcommand{\Tufts}{Physics Department, Tufts University, Medford, Massachusetts 02155, USA}
\newcommand{\WM}{Department of Physics, College of William \& Mary, Williamsburg, Virginia 23187, USA}
\newcommand{\FNAL}{Fermi National Accelerator Laboratory, Batavia, Illinois 60510, USA}
\newcommand{\Purdue}{Department of Chemistry and Physics, Purdue University Calumet, Hammond, Indiana 46323, USA}
\newcommand{\MCLA}{Massachusetts College of Liberal Arts, 375 Church Street, North Adams, MA 01247}
\newcommand{\UMD}{Department of Physics, University of Minnesota -- Duluth, Duluth, Minnesota 55812, USA}
\newcommand{\Northwestern}{Northwestern University, Evanston, Illinois 60208}
\newcommand{\UNI}{Universidad Nacional de Ingenier\'{i}a, Apartado 31139, Lima, Per\'{u}}
\newcommand{\Rochester}{University of Rochester, Rochester, New York 14627 USA}
\newcommand{\Austin}{Department of Physics, University of Texas, 1 University Station, Austin, Texas 78712, USA}
\newcommand{\USM}{Departamento de F\'{i}sica, Universidad T\'{e}cnica Federico Santa Mar\'{i}a, Avenida Espa\~{n}a 1680 Casilla 110-V, Valpara\'{i}so, Chile}
\newcommand{\Geneva}{University of Geneva, 1211 Geneva 4, Switzerland}
\newcommand{\Chicago}{Enrico Fermi Institute, University of Chicago, Chicago, IL 60637 USA}
\newcommand{\hired}{}
\newcommand{\OregonState}{Department of Physics, Oregon State University, Corvallis, Oregon 97331, USA}
\newcommand{\oxford}{}
\newcommand{\higueraThanks}{now at University of Houston, Houston, TX 77204, USA}
\newcommand{\damartinezThanks}{now at Illinois Institute of Technology, Chicago, IL 60616, USA}
\newcommand{\mcgivernThanks}{now at Iowa State University, Ames, IA 50011, USA}
\newcommand{\joelmousseauThanks}{now at University of Michigan, Ann Arbor, MI 48109, USA}
\newcommand{\twaltonThanks}{now at Fermi National Accelerator Laboratory, Batavia, IL 60510, USA}
\newcommand{\jwolcottThanks}{now at Tufts University, Medford, MA 02155, USA}

% 60 total signatories.

\author{L.~Aliaga}                        \affiliation{\WM} 
\author{M.~Kordosky}                      \affiliation{\WM} \email{makordosky@wm.edu}
\author{T.~Golan}                         \affiliation{\Rochester}  \affiliation{\FNAL}
\author{O.~Altinok}                       \affiliation{\Tufts}
\author{L.~Bellantoni}                    \affiliation{\FNAL}
\author{A.~Bercellie}                     \affiliation{\Rochester}
\author{M.~Betancourt}                    \affiliation{\FNAL}
\author{A.~Bravar}                        \affiliation{\Geneva}
\author{H.~Budd}                          \affiliation{\Rochester}
\author{M.F.~Carneiro}                    \affiliation{\CBPF}
\author{G.A.~D\'{i}az~}                   \affiliation{\Rochester}  \affiliation{\PUCP}
\author{E.~Endress}                       \affiliation{\PUCP}
\author{J.~Felix}                         \affiliation{\Guanajuato}
\author{L.~Fields}                        \affiliation{\FNAL}  \affiliation{\Northwestern}
\author{R.~Fine}                          \affiliation{\Rochester}
\author{A.M.~Gago}                        \affiliation{\PUCP}
\author{R.Galindo}                        \affiliation{\USM}
\author{H.~Gallagher}                     \affiliation{\Tufts}
\author{R.~Gran}                          \affiliation{\UMD}
\author{D.A.~Harris}                      \affiliation{\FNAL}
\author{A.~Higuera}\thanks{\higueraThanks}  \affiliation{\Rochester}  \affiliation{\Guanajuato}
\author{K.~Hurtado}                       \affiliation{\CBPF}  \affiliation{\UNI}
\author{M.~Kiveni}                        \affiliation{\FNAL}
\author{J.~Kleykamp}                      \affiliation{\Rochester}
\author{T.~Le}                            \affiliation{\Tufts}  \affiliation{\Rutgers}
\author{E.~Maher}                         \affiliation{\MCLA}
\author{W.A.~Mann}                        \affiliation{\Tufts}
\author{C.M.~Marshall}                    \affiliation{\Rochester}
\author{D.A.~Martinez~Caicedo}\thanks{\damartinezThanks}  \affiliation{\CBPF}
\author{K.S.~McFarland}                   \affiliation{\Rochester}  \affiliation{\FNAL}
\author{C.L.~McGivern}\thanks{\mcgivernThanks}  \affiliation{\Pittsburgh}
\author{A.M.~McGowan}                     \affiliation{\Rochester}
\author{B.~Messerly}                      \affiliation{\Pittsburgh}
\author{J.~Miller}                        \affiliation{\USM}
\author{A.~Mislivec}                      \affiliation{\Rochester}
\author{J.G.~Morf\'{i}n}                  \affiliation{\FNAL}
\author{J.~Mousseau}\thanks{\joelmousseauThanks}  \affiliation{\Florida}
\author{D.~Naples}                        \affiliation{\Pittsburgh}
\author{J.K.~Nelson}                      \affiliation{\WM}
\author{A.~Norrick}                       \affiliation{\WM}
\author{Nuruzzaman}                       \affiliation{\Rutgers}  \affiliation{\USM}
\author{V.~Paolone}                       \affiliation{\Pittsburgh}
\author{J.~Park}                          \affiliation{\Rochester}
\author{C.E.~Patrick}                     \affiliation{\Northwestern}
\author{G.N.~Perdue}                      \affiliation{\FNAL}  \affiliation{\Rochester}
\author{R.D.~Ransome}                     \affiliation{\Rutgers}
\author{H.~Ray}                           \affiliation{\Florida}
\author{L.~Ren}                           \affiliation{\Pittsburgh}
\author{D.~Rimal}                         \affiliation{\Florida}
\author{P.A.~Rodrigues}                   \affiliation{\Rochester}
\author{D.~Ruterbories}                   \affiliation{\Rochester}
\author{H.~Schellman}                     \affiliation{\OregonState}  \affiliation{\Northwestern}
\author{C.J.~Solano~Salinas}              \affiliation{\UNI}
\author{S.~S\'{a}nchez~Falero}            \affiliation{\PUCP}
\author{B.G.~Tice}                        \affiliation{\Rutgers}
\author{E.~Valencia}                      \affiliation{\Guanajuato}
\author{T.~Walton}\thanks{\twaltonThanks}  \affiliation{\Hampton}
\author{J.~Wolcott}\thanks{\jwolcottThanks}  \affiliation{\Rochester}
\author{M.Wospakrik}                      \affiliation{\Florida}
\author{D.~Zhang}                         \affiliation{\WM}

\collaboration{The MINER$\nu$A Collaboration}\ \noaffiliation

%% file: intro.tex
\section{Introduction} 

Precise knowledge of the neutrino flux created by an accelerator is important for precision neutrino oscillation and interaction experiments.  Conventional neutrino beams, such as the Neutrinos at the Main Injector (NuMI) beam at Fermilab, are created by directing high energy protons onto a target (usually made of carbon or beryllium) so as to produce $\pi$ and $K$ mesons. Those mesons are magnetically focused into a long tunnel in which they decay (for example, $\pi^+ \to \numu \mu^+$), producing neutrinos. In principle, precise knowledge of $\pi$ and $K$ production cross sections on the target material, and of the focusing properties of the beamline, should translate into a well known neutrino flux. In practice the situation is more complicated, since there are often multiple interactions within the target, and in the materials downstream of it. Also, the meson production process is governed by non-perturbative QCD and occurs in a nucleus, so highly accurate, first principle, theoretical predictions are not available.  Neutrino experiments have usually dealt with this situation by producing detailed simulations of the beamline materials and geometry coupled with phenomenological models of hadronic cascades, such as those in Geant4~\cite{geant4} and FLUKA~\cite{fluka1,fluka2}. Those models are not necessarily accurate but can be tuned or benchmarked by comparing their predictions to measurements of hadron production. Recent measurements of pion production on a thick (two interaction length) carbon target have been released by MIPP~\cite{mippthick}, and measurements of pion production on a thin (few per cent interaction length) carbon target are available from NA49~\cite{na49pcpi}.  In addition, there are several other hadron production measurements on various materials, using both proton and pion beams, that can be used to constrain a neutrino beamline simulation.  

%Neutrino experiments have historically relied on {\em a priori} predictions of neutrino fluxes that come from detailed simulations that incorporate a hadronic cascade model to predict what happens when high energy protons hit a target, and then account for the physical and electromagnetic properties of the rest of the beamline. 

%Thanks to the wealth of hadron production data that has become available in recent years, models that are implemented in neutrino beamline simulations can be modified to reproduce those data. 

It is also possible to directly measure the flux of mesons and muons in the beamline, thereby constraining the neutrino flux. Those measurements require the capability to count and measure the energy of the roughly $10^{6}$ particles/cm$^2$ in each beam pulse. Although measurements of these particles have been made in the past~\cite{sacha}, including one on the NuMI beamline~\cite{laurathesis}, they tend to suffer from poorly-constrained backgrounds and detector uncertainties and, at best, have achieved an accuracy of 15\%.

This article presents flux predictions based on a simulation that has been modified to reproduce  thin and thick target measurements of meson and nucleon production as well as measurements of meson and nucleon absorption cross sections.  These predictions for the $\nu_\mu$ and total neutrino fluxes are then compared to two {\em in situ} measurements that use neutrino interactions in the MINERvA detector, located 1~km from the NuMI pion production target. The two measurements use $\nu e^- \to \nu e^-$ scattering~\cite{jaewonthesis} and $\nu_\mu$ charged current ``low-$\nu$'' scattering~\cite{original_low_nu,sanjib,precision_measurements,minos_cc_xsec,ariepaper}.

%
%Importance, historical perspective, related efforts (T2K, monitor analysis/R\&D).  Strategic overview, what %we are reporting in this paper.
%\section*{NuMI beamline simulation}
%\begin{itemize}
%\item Brief description of the NuMI beam and a citation to the NuMI beam paper. Probably no figure.
%\item Description of the MC. The geometry and physics list. Citation to geant.
%\item What we record in the ntuple.
%\item Interactions map figure. 
%\item xF,pT spectrum
%\end{itemize}
%

\section{The NuMI Beam}

The NuMI beam is a wide-band neutrino beam made by impinging \gev{120} protons from Fermilab's Main Injector onto a two interaction length graphite target~\cite{numinim}. The produced pions and kaons are focused by two magnetic horns~\cite{horns} downstream of the target, each 3~m long. This reduces the charged meson angular spread, allowing them to travel out of the target hall and into a helium-filled, \unit[675]{m} long iron-walled decay pipe that has a \unit[2]{m} diameter. Neutrinos are produced when the mesons decay in flight.

From March 2005 to June 2012 the NuMI beamline operated primarily in the ``low energy'' (LE) configuration.  In this configuration the downstream end of the 95~cm long target was inserted 57~cm past the front face of the first horn, and both horns (separated by 10~m) were pulsed at 185~kA~\cite{numinim}.  
%, optimized for MINOS's $\numu \to \numu$ and $\numu \to \nue$ oscillation searches. 
%The target in the LE configuration was able to be moved along the beamline, changing the beam
% focusing and energy spectrum. The nominal target postion has it fully inserted into the first horn.
% However, almost all the data relevant for MINOS oscillation and MINERvA cross section analyses were
% taken with the target located upstream of its nominal position by 10~cm and with a horn current of
%The horns in this configuration are pulsed at 185~kA and are separated from eachother by 10 meters.
In this configuration, the peak neutrino energy is \gev{3.5} with a high-energy tail extending to several tens of GeV.  
When the horn current is set to focus positive particles the resulting beam consists of 93\% $\nu_\mu$, 6\% $\bar\nu_\mu$ and 1\% $\nu_e+\bar\nu_e$. This was the configuration that accumulated the most protons on target (POT) during the period defined above.  The horn current can also be set to focus negative particles to enhance the \numubar composition of the beam, and that was the configuration with the next largest accumulated number of protons on target. 
Small (few-week) datasets were taken with the target pulled back from the horn by 100-250~cm, creating higher energy beams used for systematic studies~\cite{Zwaska,laurathesis}. 
%The low energy NuMI beam is described in greater detail in~\cite{numinim}.

NuMI is simulated using Geant4~\cite{geant4}\footnote{Geant v4.9.2.p03 was used with the FTFP-BERT physics list.} and a detailed geometrical model of the beamline, which was originally created for MINOS~\cite{zarkothesis,laurathesis,jasonthesis} and subsequently improved by MINERvA~\cite{leothesis}. The simulation accounts for all particle interactions and propagation in the beamline, starting with protons entering the carbon target and ending in decays that produce a neutrino. The effect of target aging due to radiation damage does not appear to be a significant effect during the period in which MINERvA took data and is not simulated.
  
The simulation outputs the location and kinematic information of each decay producing a neutrino. The neutrino flux at a particular location is then determined by using the differential decay rate, as a function of solid angle, given the neutrino species and the parent particle's kinematic information. 
%
%We record the entire chain of interactions  particles in 
% v9.2.p3
%Uncertainties on the neutrino flux come from two main sources: (1) the material makeup, shape and %alignment of beamline elements and (2) hadronic interactions. The both of these are 

%% file: hadroproduction.tex
\section{Hadronic Interactions in NuMI} 

In this section we describe the processes that produce neutrinos in the NuMI beamline, identify what measurements can constrain these processes, and evaluate the associated uncertainties.  The interactions that occur in the NuMI beamline can first be categorized by incident particle and target material.  Roughly 85\% of the interactions that produce particles that lead to muon neutrinos passing through MINERvA are from protons interacting on carbon. Other relevant materials are aluminum (horns), iron (decay pipe walls), helium (decay pipe gas), and air (target hall). Interactions of $\pi^\pm$, $K^\pm$ and $n$ created in the initial proton interaction, or subsequent interactions, are subdominant but non-negligible. Table~\ref{tab:interactions} summarizes the hadronic interactions that lead to \numu that pass through MINERvA.

\input{table1.tex}

\begin{figure}
\centering
\includegraphics[width=\columnwidth]{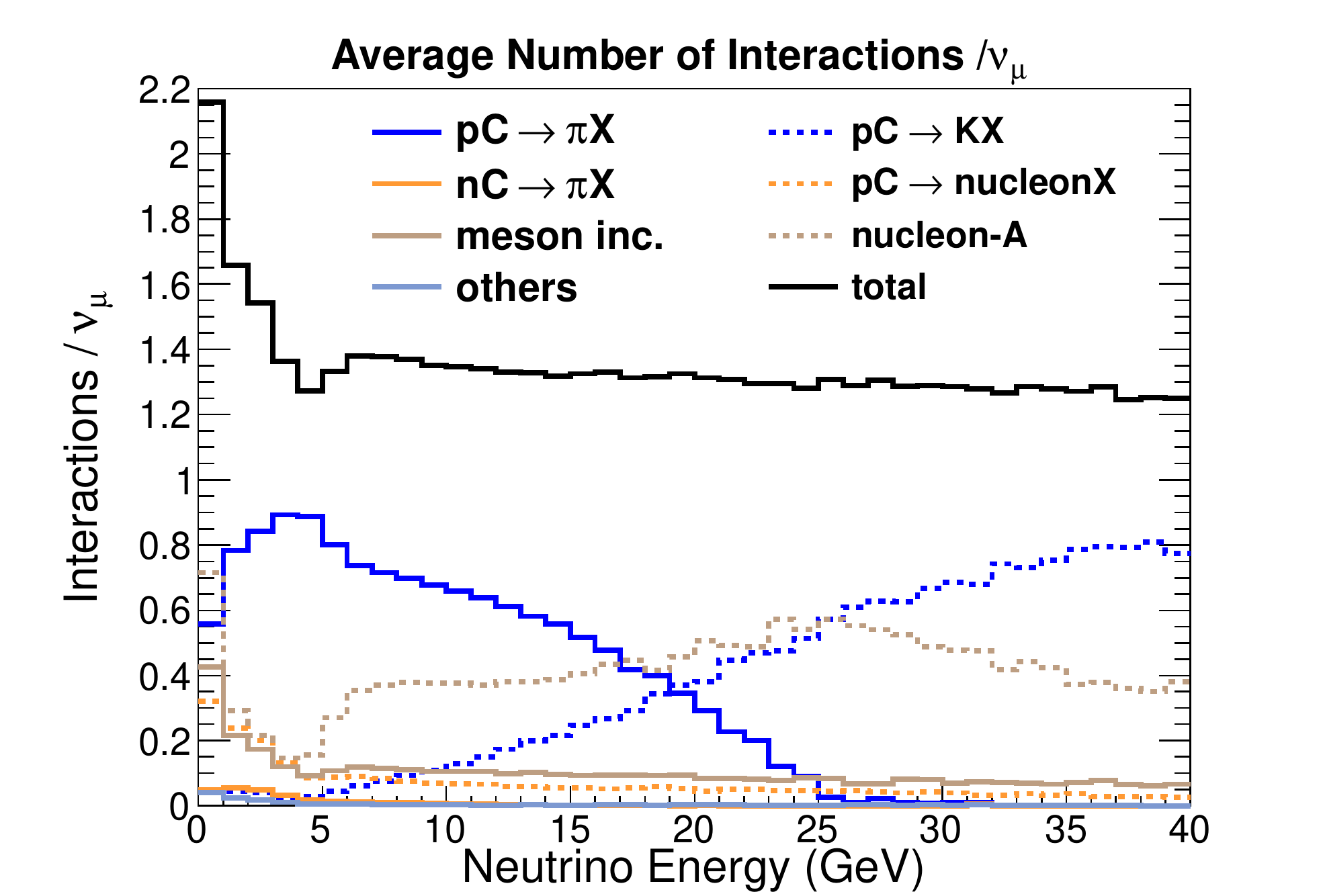}
\caption{The number of interactions per \numu passing through MINERvA as a function of the neutrino energy in the LE beam configuration.  The lines show the different categories of interactions for which we apply constraints and uncertainties based on hadron production data:  ``nucleon-A'' ~refers mainly to nucleons interacting in material that is not carbon~\cite{nucleonA}, and ``meson inc.'' ~refers to mesons interacting on any material in the beamline.  }
\label{fig:imap}
\end{figure}

When protons collide with carbon, the interactions can produce pions, kaons, neutrons, strange baryons, and lower energy protons.  These particles, if they do not decay first, can interact either in the target or in other downstream material to create tertiary particles that can also decay into neutrinos.  Figure~\ref{fig:imap} shows the number of interactions in each of these categories, including those of the primary \unit[120]{GeV/c} protons, as a function of the produced $\nu_\mu$ energy in the NuMI beamline, for the LE configuration.  There are a small number of interactions that do not fit into any of the categories above, and are rare, only affecting the energy bins below $\gev{1}$.

%Fig.~\ref{fig:imap}. In this section we will describe each sub-category, including the data and/or %theoretical ideas that we use in our constraints. 
%The end result of our work is the flux prediction in Fig.~\ref{fig:le_flux} and the error band in %Fig.~\ref{fig:le_unc}.

%\begin{description}
%
%\item[$\mathbf{pC\to \pi X}$] 
There are two major datasets available to constrain the process where protons interact on carbon and produce charged pions. One measurement, from NA49~\cite{na49pcpi}, uses a thin target with an incident proton momentum of \unit[158]{GeV/c}.  The other measurement, from MIPP~\cite{mippthick}, uses an actual NuMI LE target and \unit[120]{GeV/c} protons. These two datasets will be used to make separate ``thin target'' and ``thick target'' flux predictions by weighting each interaction leading to a neutrino going through MINERvA. We also use additional datasets to constrain kaon and nucleon production, and the absorption of particles in beamline materials. Where multiple interactions are constrained with data, the overall weight applied to the neutrino event is simply the product of the weights for each interaction.

For the thin target prediction we use NA49's measured invariant cross section for pion production~\cite{na49pcpi}, \invxs, to compute the $\pi^{\pm}$ yield per inelastic interaction, 
\begin{equation}
f_{Data}=\frac{1}{\siginel}\invxs 
\end{equation}

\begin{figure}
\centering
\includegraphics[width=\columnwidth]{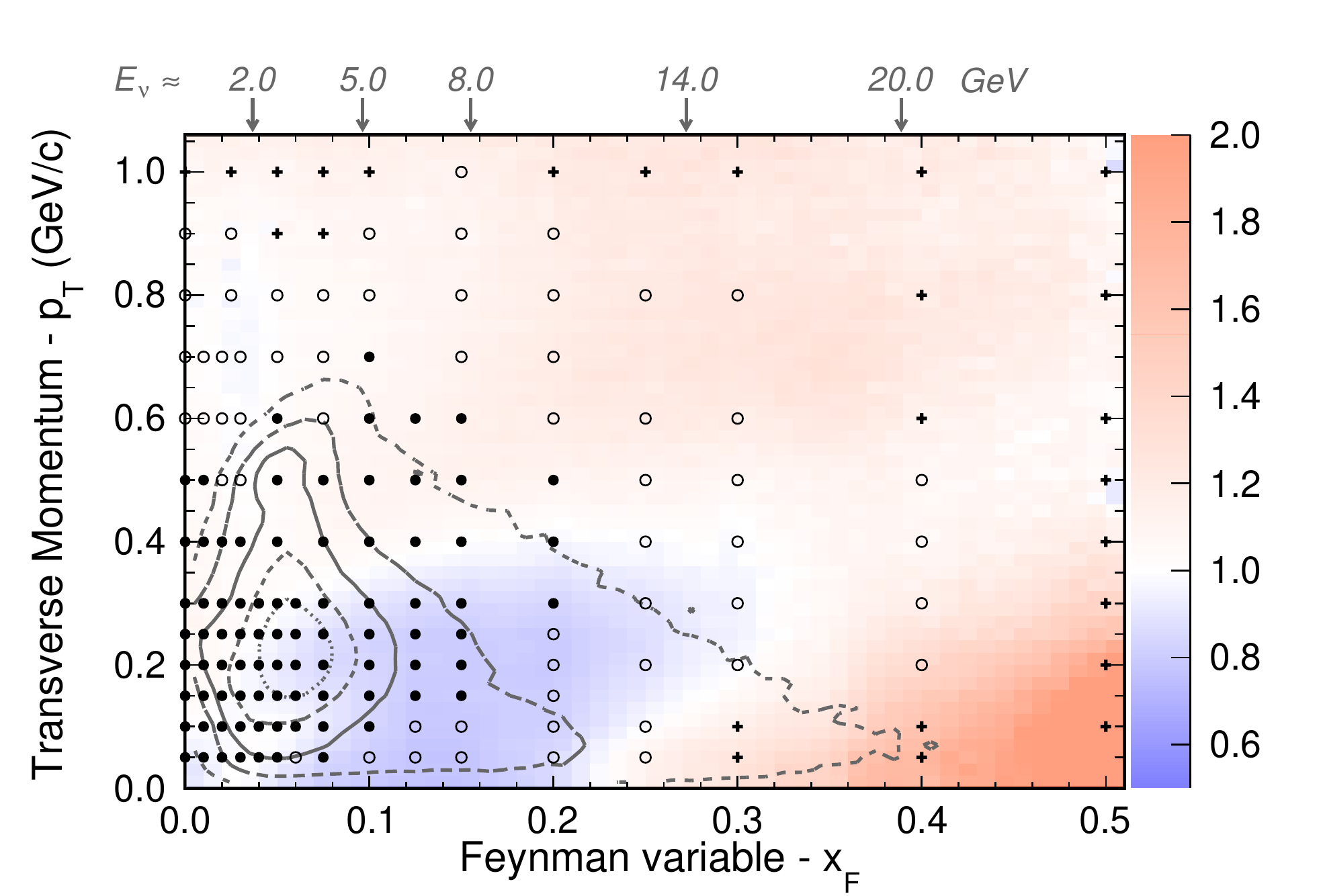}
\caption{A summary of the hadron production data from NA49 and its application in the analysis to predict the ``thin target'' flux. The markers show the location of NA49's measurements of the invariant cross section for $p\,C \to \pi^+ X$ interactions as a function of the produced pion's kinematics in (\xf,\pt) plane. The marker types correspond to statistical uncertainties $< 2.5\%$ ($\bullet$), $<5\%$ ($\circ$)  and $>5\%$ (\textbf{+}). The color scale shows the data/MC ratio $f_{Data}/f_{MC}$ applied in Eq.~\ref{eq:weight} to correct the simulation. The topographical contours indicate the number of  $p\,C \to \pi^+ X$ interactions leading to \numu in MINERvA in the LE beam. From inner to outer these are at 75, 50, 25, 10, and 2.5\% of the peak value. The upper axis shows the approximate energy of a $\nu_\mu$ produced by a pion at the corresponding \xf.
}
\label{fig:xfpt}
\end{figure}

\begin{figure}
\centering
\includegraphics[width=\columnwidth]{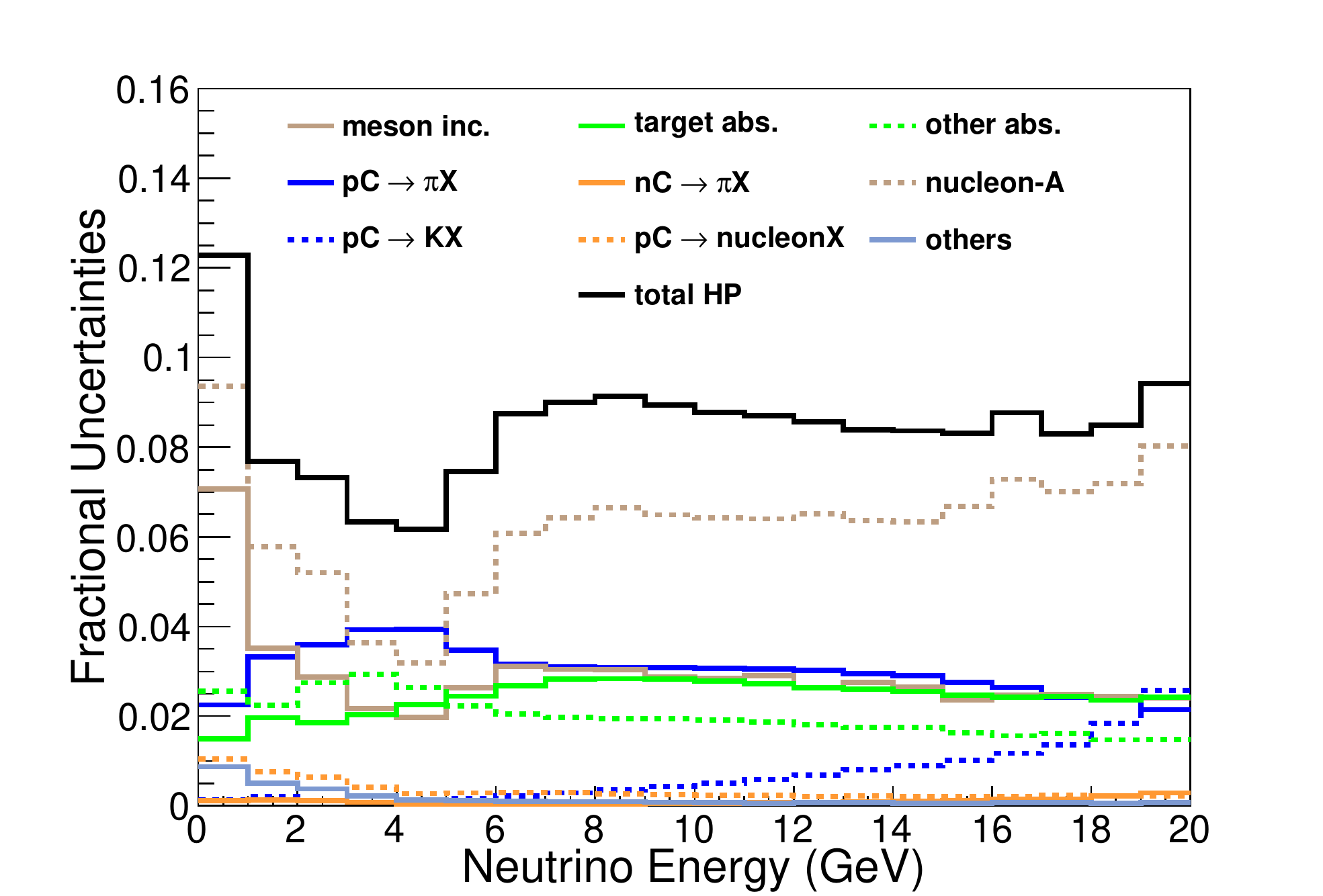}
\caption{Uncertainties on the NuMI low energy \numu  ``thin target flux'' that originate from the different hadron interaction categories described in the text.   The label ``nucleon-A'' ~refers mainly to nucleons interacting in material that is not carbon~\cite{nucleonA}, and  ``meson inc.'' ~refers to mesons interacting on any material in the beamline; ``target abs.'' and ``other abs.'' ~refer to absorption in the target (C) and other materials (Al, He, Fe).}
\label{fig:thin_unc}
\end{figure}

Here, $E_\pi$ is the energy of the pion.  The factor \siginel is inserted here to convert the invariant cross section into a yield. The impact of the uncertainty on \siginel is considered later in this paper. The prediction for the same quantity, $f_{MC}$, is used to produce weights that we apply to the simulated pion production yield to bring the simulation into agreement with the data:  
\begin{eqnarray} \label{eq:weight}
w(x_F,p_T,p) &=& \frac{f_{Data}(x_F,p_T,p_{0}=\gevc{158})}{f_{MC}(x_F,p_T,p_{0}=\gevc{158})}  \times s(x_F,p_T,p)
\end{eqnarray}
The cross sections and weights are functions of the proton's momentum $p$, the Feynman variable \xf and the transverse momentum \pt.  NA49 quotes a systematic uncertainty of 3.8\% that we assume is 100\% bin-to-bin correlated in the error propagation procedure described later in this paper~\cite{na49_systematic}. Motivated by Feynman scaling~\cite{feynmanscaling} we also apply a scale ($s$) to translate from \gevc{158} to proton momenta between 12 and \gevc{120} using FLUKA~\cite{fluka1,fluka2}:
\begin{eqnarray} 
s(x_F,p_T,p) &=& \frac{\sigma_{FLUKA}(x_F,p_T,p)}{\sigma_{FLUKA}(x_F,p_T,p_{0}=\gevc{158})}
\end{eqnarray}
  This prescription was checked by scaling NA49 pion production data at \gevc{158} to NA61 data taken at \gevc{31}~\cite{na61pcpi}. The difference between the two was less than 5\% across-the (\xf,\pt) region in which both experiments took data. We propagated that difference as an uncertainty on the flux and found that it was negligible~\cite{jeremy_thesis}. Figure~\ref{fig:xfpt} shows the statistical uncertainties on the NA49 pion production data and the amount by which the standard simulation must be weighted in order to achieve agreement with that data set. The neutrinos at the peak of the NuMI beam preferentially come from the highest statistics center of the NA49 data set, where the center contour of Fig.~\ref{fig:xfpt} overlaps high precision data points. This translates into a relatively small flux uncertainty in the few-GeV neutrino energy range, as shown in Fig.~\ref{fig:thin_unc}.

We apply weights from the NA49 data for $\xf<0.5$ and use the dataset of Barton {\it et al.}~\cite{barton} for $0.5<\xf<0.88$ and $0.3<\pt<\unit[0.5]{GeV/c}$. The Barton and NA49 datasets disagree by approximately 25\% where they overlap, while the uncertainties on each are only a few percent. We normalize Barton to NA49 in the overlap region and assign a 25\% uncertainty to all of the Barton data. We then construct a bin-to-bin covariance matrix between the various $(\xf,\pt)$ bins.  In the remainder of this paper we will refer to the flux constrained using these data, as well as the other thin target datasets described later, as the ``thin target flux''.  

%$\item[$\mathbf{pC\to K X}$]  

\begin{figure}
\centering
\includegraphics[width=\columnwidth]{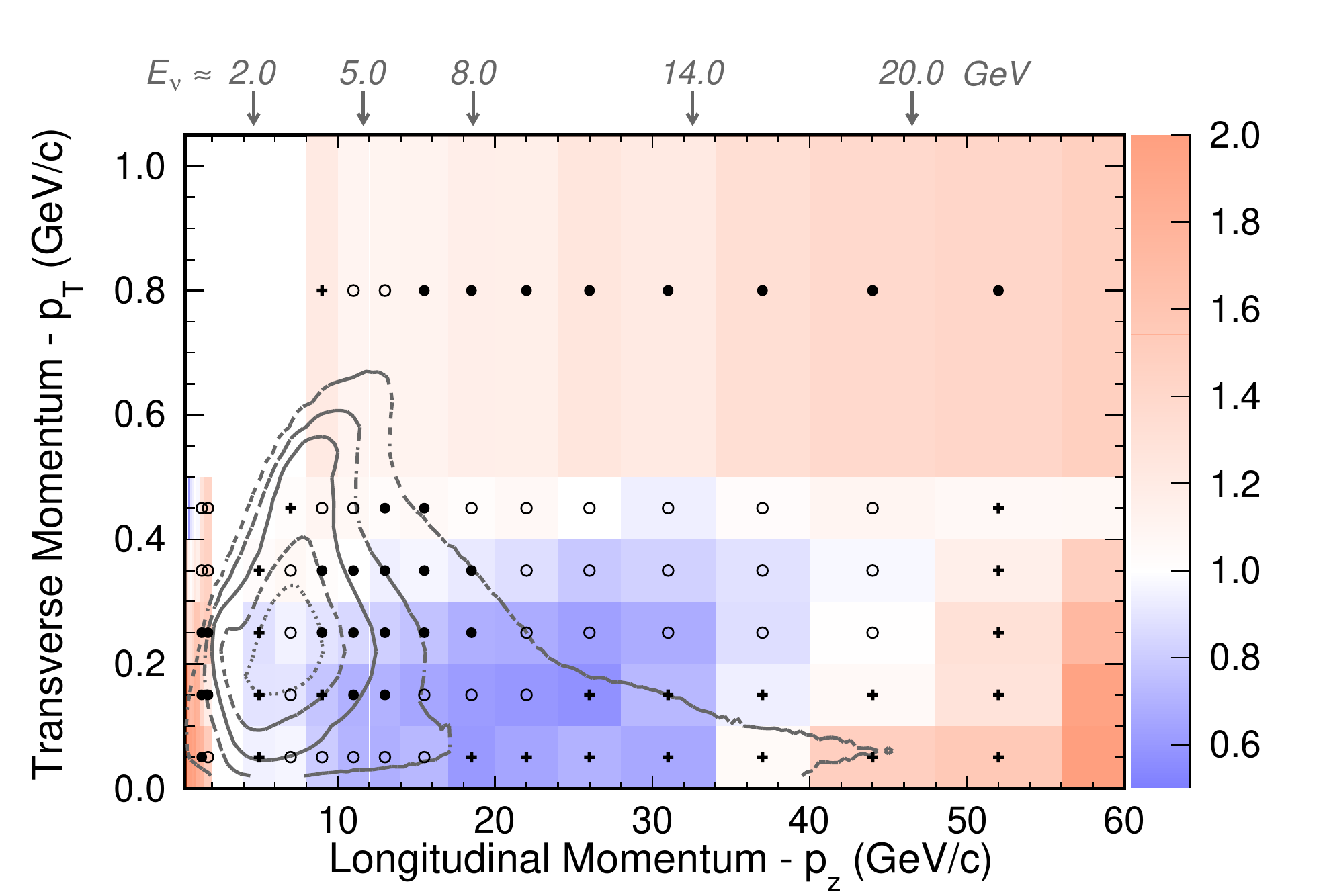}
\caption{A summary of the hadron production data from MIPP and its application to predict the ``thick target'' flux. The markers show the bin-center of MIPP's measurements of $\pi^+$ yields from \gev{120} protons interacting with a NuMI target. The upper \pt bins extend to \gevc{2} but the markers are drawn at \gevc{0.8}. The marker types correspond to statistical uncertainties $< 2.5\%$ ($\bullet$), $<5\%$ ($\circ$)  and $>5\%$ (\textbf{+}). The color scale shows the data/MC ratio applied to correct the simulation. The topographical contours indicate the number of $\pi^+$ exiting the target that lead to \numu in MINERvA in the LE beam. From inner to outer these are at 75, 50, 25, 10, and 2.5\% of the peak value. The upper axis shows the approximate energy of a $\nu_\mu$ produced by a pion at the corresponding \pz.
}
\label{fig:mipp}
\end{figure}

The MIPP thick target yields cover most of the region $1<\pz<\gevc{80}$, $0<\pt<\gevc{2}$. We use these data by tabulating pions leaving the simulated target as a function of \xf and \pt. Each is then weighted by the ratio of the yield measured by MIPP and the yield predicted by the simulation. These weights account for pions produced by the original proton and also for reinteractions in the target. The MIPP statistical uncertainties range from approximately 2-6\% in the kinematic bins of interest and have a roughly 5\% systematic uncertainty that we assume is 75\% correlated, bin-to-bin~\cite{mipp_systematic}. Figure~\ref{fig:mipp} shows the statistical uncertainties on the MIPP pion production data and the amount by which the standard simulation must be weighted in order to achieve agreement with that data set.

The $K/\pi$ production ratio from the NuMI target was also measured in the region $20<\pz<\unit[90]{GeV/c}$, $\pt<\unit[1]{GeV}$~\cite{sharonthesis}. The ratio is used with the pion yields to estimate the kaon yields. The data have statistical uncertainties generally in the 5-20\% range and systematic uncertainties in the several percent range, which are added in quadrature.  Hereafter we will refer to the flux constrained using the MIPP NuMI target data as the ``thick target flux''.

There are several other datasets that are used in the thin target flux prediction. These data are also used to fill in gaps in the thick target data when making the thick target  prediction. NA49 measured $p\,C\rightarrow K^\pm X$ for $0\leq \xf \leq 0.2$, $0.1\leq \pt \leq 0.9$~GeV~\cite{gemmathesis}. The uncertainty is dominated by statistics, so the combination of statistical and systematic uncertainties (approximately 5-10\%) is applied assuming no bin-to-bin correlations. For $0.2<\xf<0.5$ we utilize the ratio of $K/\pi$ yields on a thin carbon target as measured by MIPP~\cite{andrethesis}, multiplied by the NA49 thin target yields described above. The uncertainty on the $K/\pi$ data is large, typically $\cal{O}$(10\%), and dominated by the subtraction of large pion and proton backgrounds when the kaon yields are estimated. We assume the uncertainties are uncorrelated from one bin to the next and do not incorporate the relatively small uncertainty contribution from NA49 pion yields that appear in the denominator of the $K/\pi$ ratio.   

%\item[$\mathbf{pC\to nucleon X}$] 
Nucleon production in $p\,C$ collisions is constrained using data from NA49~\cite{na49_nucleon}. The data cover $-0.8 \leq \xf \leq 0.95$, $0.05 \leq \pt \leq \gevc{1.9}$ for produced protons. For neutrons the data are integrated over \pt and cover $0.1 \leq \xf \leq 0.9$. Both datasets have statistical uncertainties $\lesssim 10\%$ except in the most extreme bins.  Systematic uncertainties are 3.7\% for protons and 10\% for neutrons. We assume the systematics are 100\% bin-to-bin correlated. Weights are derived using the same procedure we used for pion and kaon production.

%\item[$\mathbf{nC \to \pi X}$] \
Neutron induced pion production off of carbon is constrained by extending isospin symmetry in reactions with a deuterium (isoscalar) target, $\sigma(p d \to \pi^+ X d)=\sigma(n d \to \pi^- p X)$, to carbon, treating $p\,C \to \pi^+ X$ data as $n\,C \to \pi^- X$ and vice versa. Neutron interactions make a small contribution to the neutrino flux, as shown in Fig.~\ref{fig:imap}.  We assume the uncertainty is that of the $p\,C$ data, as no relevant data to test this {\it ansatz} exists and the correction's impact is small.

%\item[nucleon-A] 
A subdominant portion of the flux involves nucleon interactions on nuclei that are not carbon, most commonly He, Fe and Al. We constrain these interactions with thin target $p\,C$ data whenever the produced particles are within the kinematic range of that data. The additional uncertainty due to this procedure was estimated as follows. First measurements of $K^0,\Lambda^0$, and  $\bar{\Lambda^0}$ production off of Be, Cu and Pb targets by a \gev{300} proton beam~\cite{skubic} are used to derive an $A$-dependent scaling~\cite{barton,skubic} in bins of momentum and angle. This scaling was then applied to the simulation and tested against measurements of $p\,A\to \pi X$ and $p\,A \to K X$ data collected at \gev{100} on C, Al, Cu, Ag, and Pb targets~\cite{barton}. Discrepancies between the predicted and measured yields are incorporated as uncertainties. These discrepancies range from 2.5 to 30\%, depending on the produced particle and the kinematic bin.

%\item[Meson incident] 
% So, we are forced to make an {\it ab initio} guess at the uncertainties.

Mesons traversing beamline elements often interact to produce particles that eventually lead to a neutrino. Unfortunately there is little applicable data for the 10-\gev{40} mesons of interest here. We estimate the uncertainty by noting that Geant4-FTFP is a microphysical, first principles model of hadronic interactions. Our {\it ansatz} is that the level of agreement between FTFP and existing hadron production datasets is indicative of FTFP's ability to model interactions for which no data is currently available.  Meson and nucleon production measurements exist for $p\,C$ and, more generally, for $p\,A$ interactions. That data agrees with the simulation at better than 40\% across a broad range of relevant kinematics.  We assume that this verifies the FTFP model at the 40\% level. In addition, we note that the observed data-simulation discrepancies for production of $\pi^\pm, K^\pm, n$ and $p$ do not appear to be correlated in any obvious way. Therefore, to handle meson incident interactions we categorize the interactions based on incident particle $(\pi^\pm, K^\pm)$ and produced particle  $(\pi^\pm, K^\pm, n, p)$. For each combination we break the range $0<\xf<1$ into 4 equally sized bins. In each bin we assign a 40\% uncertainty and we treat each bin as being uncorrelated with the others.

%For each we assume a  40\% uncorrelated uncertainty in 4 \xf bins, equally spaced in the range $0<\xf<1$.

Sometimes nucleons interact and produce particles that are outside the kinematic coverage of any dataset. We categorize these interactions in terms of incident particle ($n,p$) and produced particle ($n,p,\pi^\pm,K^\pm$). As for incident mesons, we assume a 40\% uncorrelated uncertainty in 4 \xf bins, equally spaced in the range $0<\xf<1$. In this category of interactions, the largest contributor to the overall flux uncertainty comes from quasi-elastic $p\,C\to pX$ interactions (defined as nucleon knockout without mesons or heavy baryons) at $\xf>0.95$.

Any interactions not covered above are combined in an ``other'' category and assigned a single 40\% uncertainty.  This is consistent with the uncertainties assigned to other Geant4-FTFP predictions.

%\item[Others] 
%
%\section{Hadronic Attenuation and Absorption}
%

%We have described a detailed procedure for correctly constraining the simulation of hadron production, and we follow a similar procedure for hadron attenuation.  The importance of these constraints is given by Fig.~\ref{fig:material} which shows how much material is traversed by pions that produce \numu's that reach the MINERvA detector, as a function of produced $\nu_\mu$ energy.  

\begin{figure}
\centering
\includegraphics[width=\columnwidth]{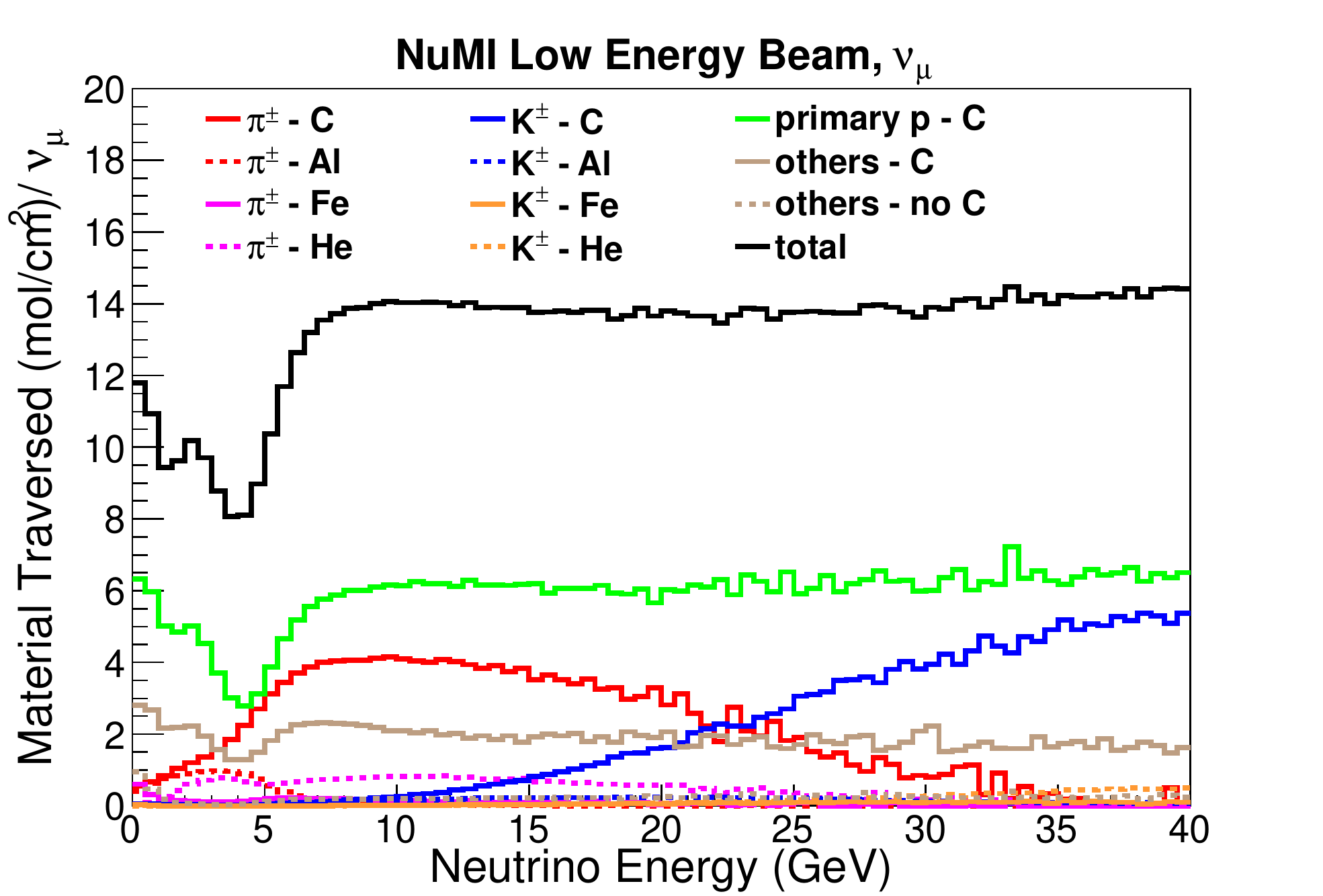}
\caption{A summary of the material traversed by \pip parents of \numu incident on MINERvA in the LE beam configuration. This material accounting is used to compute the impact that uncertainties on hadron absorption cross sections have on the flux.}
\label{fig:material}
\end{figure}

%\item[target attenuation] 
%
%taken from~\cite{bellettini}.

Particles in the NuMI beamline travel through a significant amount of material.  As shown in Fig.~\ref{fig:material}, the carbon in the target is (by design) the most frequently encountered material, with protons typically traversing $\sim \unit[6]{mol/cm^2} \approx \unit[40]{cm}$.  Pions and kaons also travel through a significant amount of carbon as they leave the target, and then later encounter aluminum in the horns, helium in the decay pipe volume, and iron in the decay pipe walls. The flux uncertainty is impacted by imperfect modeling of nucleon and meson absorption cross sections, \sigabs, in those materials. Uncertainties in \sigabs translate into uncertainties on the rate of interactions and the location of those interactions in the beamline. The position of interactions in the target is especially critical since that influences the focusing of the produced particles.

The absorption cross section is defined as the  sum of the inelastic cross section, \siginel (meson and heavy baryon production), and the quasi-elastic cross section. Several precise measurements are available to constrain \siginel in $p\,C$ collisions~\cite{denisov,carroll,na61pcpi}. The measurements show that the simulation underpredicts this cross section at the 5\% level. We correct for the discrepancy and also adopt a 5\% uncertainty. The $p\,C$ quasi-elastic cross section is taken to be $\unit[29\pm 4]{mb}$. It was computed from the average and spread of the cross-sections reported by T2K~\cite{t2kflux}, NA61~\cite{na61pcpi}, and Gaisser {\it et al}~\cite{gaisser}. We use a 40\% uncertainty for $n\,C$ interactions and also for $n$ and $p$ collisions with He, Al and Fe.  We do not correct the inelastic cross-section in any of those cases.

We propagate the uncertainty in \sigabs by tabulating the amount of material traversed by each of the particles leading to a neutrino, and recording if they were absorbed or not. The probability that a particle does not interact when crossing through a material of length $z$ is $P_\mathrm{survival}(z) = \exp(- z N_A \rho \sigabs)$, where $\rho$ is the nuclear number density.  
%The probability it is absorbed is in a thin layer $\Delta z$ is $P_\mathrm{abs}(z) = \Delta Z N_A \rho \sigabs \exp(- z N_A \rho \sigabs)$. 
For C, He, Al, and Fe we compute the appropriate probabilities using \sigabs from both the data and simulation and assign the ratio as a weight. Absorption from other materials is negligible.  In the thin target prediction the way in which NA49's measurement depends on \siginel was removed in the computation of $f_{Data}$ so as to avoid double counting that uncertainty at this stage. For the thick target prediction, the initial $p\,C$ interaction in the target is reweighted according to the formula above, but a correction is made to assure that the average weight does not deviate from unity. This avoids altering the yields, since that would double count the uncertainty already incorporated in the MIPP uncertainties, but allows the average position of interactions along the target to vary.

Absorption uncertainties for pions and kaons are handled in a similar way. The simulation reproduces $\pi^\pm C$ and $\pi^\pm Al$ datasets~\cite{cronin,denisov,allaby,longo,bobchenko,fedorov,carroll} to within 5\% for pion momenta ranging from \gev{1} to \gev{60}, so we adopt a 5\% uncertainty for $\pi A$ absorption. The $K^\pm C$ and $K^\pm Al$ cross sections measured by~\cite{abrams,denisov,carroll,allaby} are less well reproduced by the simulation. The adopted uncertainties range from 30\% at low energy ($p<\gevc{2}$) to about 10\% at high energy ($p\approx \gevc{50}$). No correction to the cross section is done for either pions or kaons, we just propagate an uncertainty. When using thick target data, the effect of pion and kaon absorption in the target is captured in the thick target yields and the position effect cannot be deconvolved. Because of this we do not propagate absorption uncertainties for pions and kaons in the target material when using thick target data. 

%\end{description}
%\section{Summary of Hadronic Uncertainties}
%
%However, before going further, we need to describe how we propagate uncertainties on individual %interactions to the neutrino energy distribution. 
%~\cite{*[{Sampling from a multi-dimensional Gaussian distribution is described in chapter 7.4 of }] nr}.

\begin{figure}
\centering
\includegraphics[width=\columnwidth]{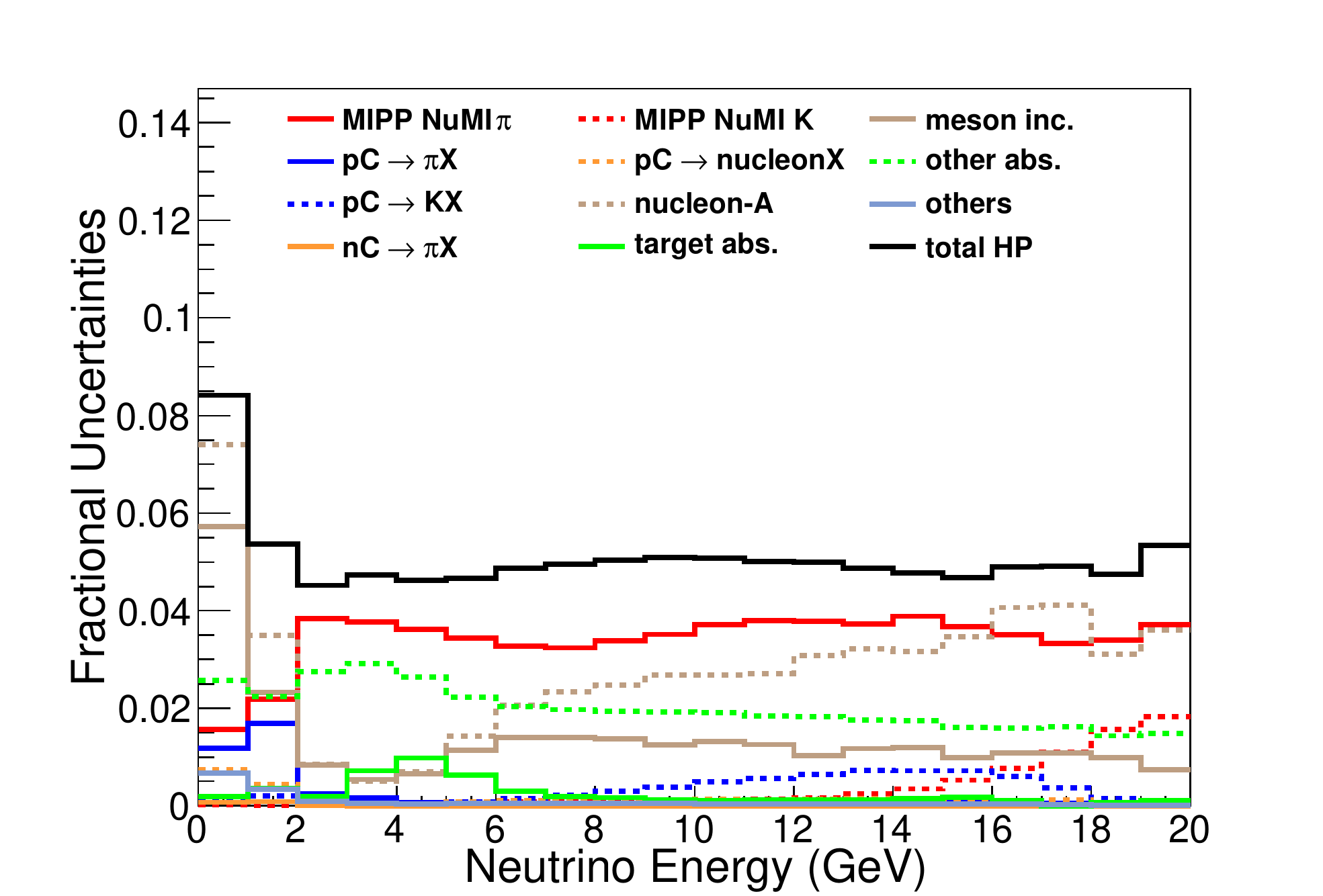}
\caption{Uncertainties on the NuMI low energy \numu ``thick target flux'' that originate from the different hadron interaction categories described in the text.   The label ``nucleon-A'' ~refers mainly to nucleons interacting in material that is not carbon~\cite{nucleonA}, and  ``meson inc.'' ~refers to mesons interacting on any material in the beamline; ``target abs.'' and ``other abs.'' ~refer to absorption in the target (C) and other materials (Al, He, Fe).}
\label{fig:thick_unc}
\end{figure}

The uncertainties described above are propagated to the neutrino energy distribution using a technique referred to as ``multi-universes''. This method envisions each hadron production data point and every other source of uncertainty listed above as a parameter with an uncertainty and possible correlations with other parameters. We construct a series of $N=100$ alternative parameter sets by randomly sampling from a multi-dimensional Gaussian distribution centered on the default parameter values with covariances to account for the uncertainties and correlations~\cite{nr}. The resulting $N$ flux predictions are used to compute the variance in each neutrino energy bin and the covariance between bins.  Figures~\ref{fig:thin_unc} and \ref{fig:thick_unc} show the resulting flux uncertainties as a function of \numu energy.

%% file: table1.tex
\begin{table}[]
\centering
\begin{tabular}{|c|d|d|d|d|d|d|d|}
\hline
\multicolumn{8}{|c|}{\textbf{Material}} \\  \hline 
\textbf{Projectile} &\mathrm{C} & \mathrm{Fe} & \mathrm{Al} & \mathrm{Air} & \mathrm{He} & \mathrm{H_2 O} & \mathrm{Be}  \\
\hline
p & 117.5 & 2.9 & 1.0 & 1.1 & 1.5 & 0.1 & 0.1 \\ 
$\pi^+$ & 8.1 & 1.3 & 1.8 & 0.2 & - & 0.4 & - \\
$\pi^-$ & 1.3 & 0.2 & 0.2 & - & - & - & - \\
$K^\pm$ & 0.6 & 0.1 & 0.1 & - & - & - & - \\
$K^0$ & 0.6 & - & - & - & - & - & - \\
$\Lambda/\Sigma$ & 1.0 & - & - & - & - & - & - \\
\hline
\end{tabular}
\caption{The average number of interactions leading to a \numu in the \minerva detector with $0<E_\nu<\gev{20}$. The numbers have been multiplied by 100 for clarity. For example, there are 1.175 pC interactions and 0.081 $\pi^+\mathrm{C}$ interactions per \numu, indicating the importance of secondary interactions in the carbon target.}
\label{tab:interactions}
\end{table}

%% file: focusing.tex
\section{Uncertainties due to the Beamline Geometry} 

Once the hadrons are produced in the target they propagate through the inner conductors and magnetic fields of horns and then through the decay pipe.  There are a large number of geometric and magnetic details that can affect the neutrino energy distribution, and those details must be precisely measured and then incorporated in the neutrino beam simulation.  

Alignment tolerances on the primary proton beam trajectory, target, and horns are described in Refs.~\cite{numinim,zarkothesis,Zwaska}. The largest effects on the flux prediction come from uncertainties on the transverse position of the most upstream horn relative to the target ($\pm0.1$~cm) and the longitudinal position of the target with respect to that horn ($\pm1$~cm).  

The magnetic field is determined not only by the current (185~kA $\pm$ 1\%) but also by the precise shape of the inner conductors, in particular of the first horn. The parabolic inner conductor is modeled in Geant as a series of {\tt G4Cone} or, alternatively, {\tt G4Polycone} volumes. An uncertainty due to the modeling is assigned by comparing the flux obtained using each of the two cone types and by varying the number of cones used in the model. Finally, the horns have a water jet cooling system that results in a residual layer of water on the horn inner conductor. That $\unit[1.0\pm0.5]{mm}$ thick layer affects the number of mesons absorbed in the horns, resulting in an uncertainty on the flux.

There is a graphite baffle just upstream of the target that protects the horn inner conductors from a mis-steered proton beam. Under normal operations the beam has small non-Gaussian tails in its radial profile that interact with (``scrape'') the baffle. Measurements of the beam profile upstream of the baffle and temperature changes in the baffle indicate that the tails make up less than 0.25\% of the beam power~\cite{zarkothesis}. We conservatively adopt that as a systematic error.  There is an additional uncertainty in the flux prediction coming from the 2\% measurement uncertainty of the number of protons incident on the NuMI target (POT counting).  

\begin{figure}
\centering
\includegraphics[width=\columnwidth]{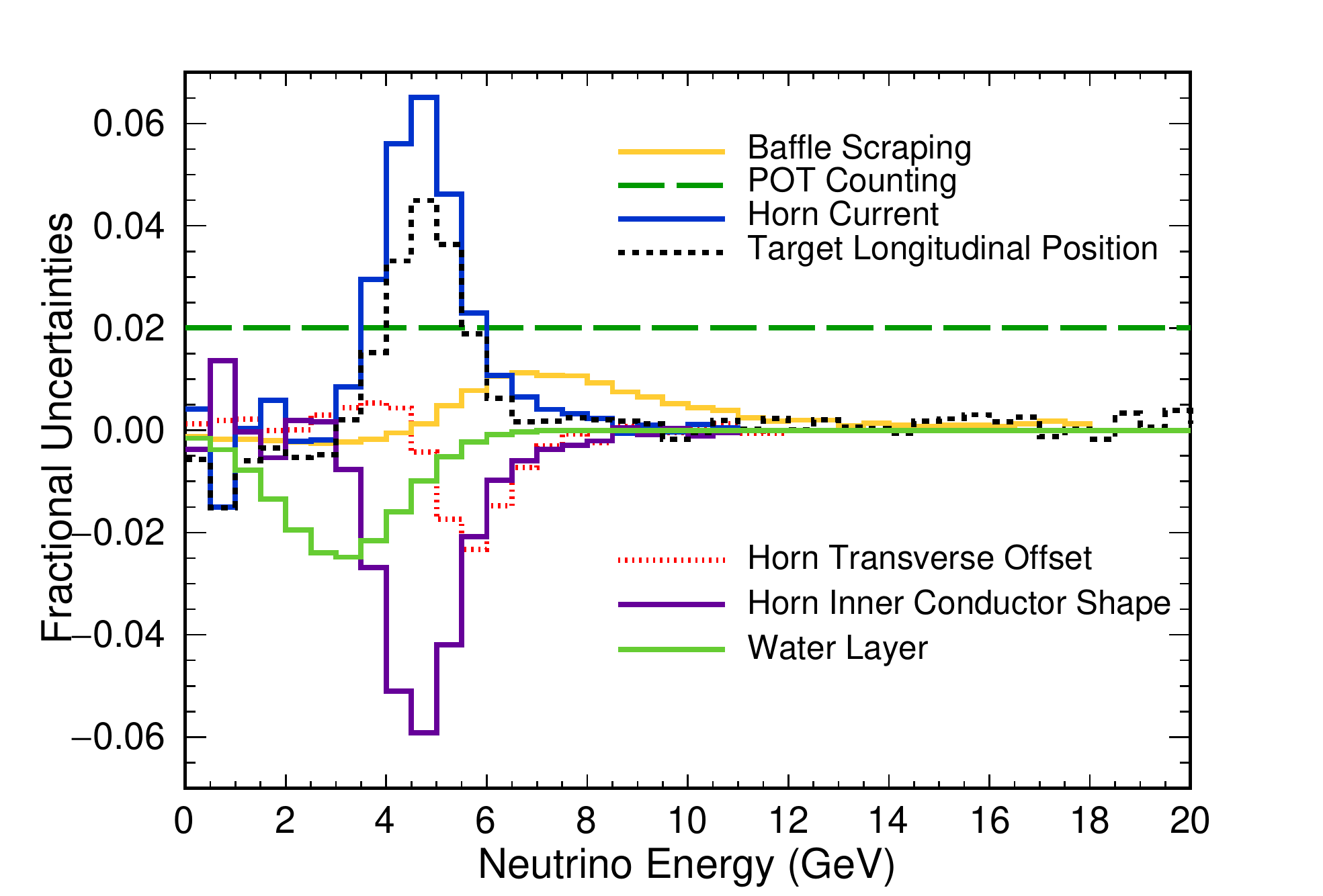}
\caption{Beam geometry and focusing uncertainties on the \numu flux at MINERvA in the LE beam configuration.  The sources of uncertainty are described in the text.}
\label{fig:foc_unc}
\end{figure}

Figure~\ref{fig:foc_unc} shows the uncertainty on the NuMI on-axis $\nu_\mu$ flux that comes from each of the focusing uncertainties.  While most of these uncertainties are smaller than those coming from the hadron production, they dominate at the $4-6$~GeV region, which is the falling edge of the neutrino energy distribution, shown in Figs.~\ref{fig:le_thin_flux} and \ref{fig:le_thick_flux}.

%% file: prediction.tex
\section{Results}

\begin{figure}
\centering
\includegraphics[width=\columnwidth]{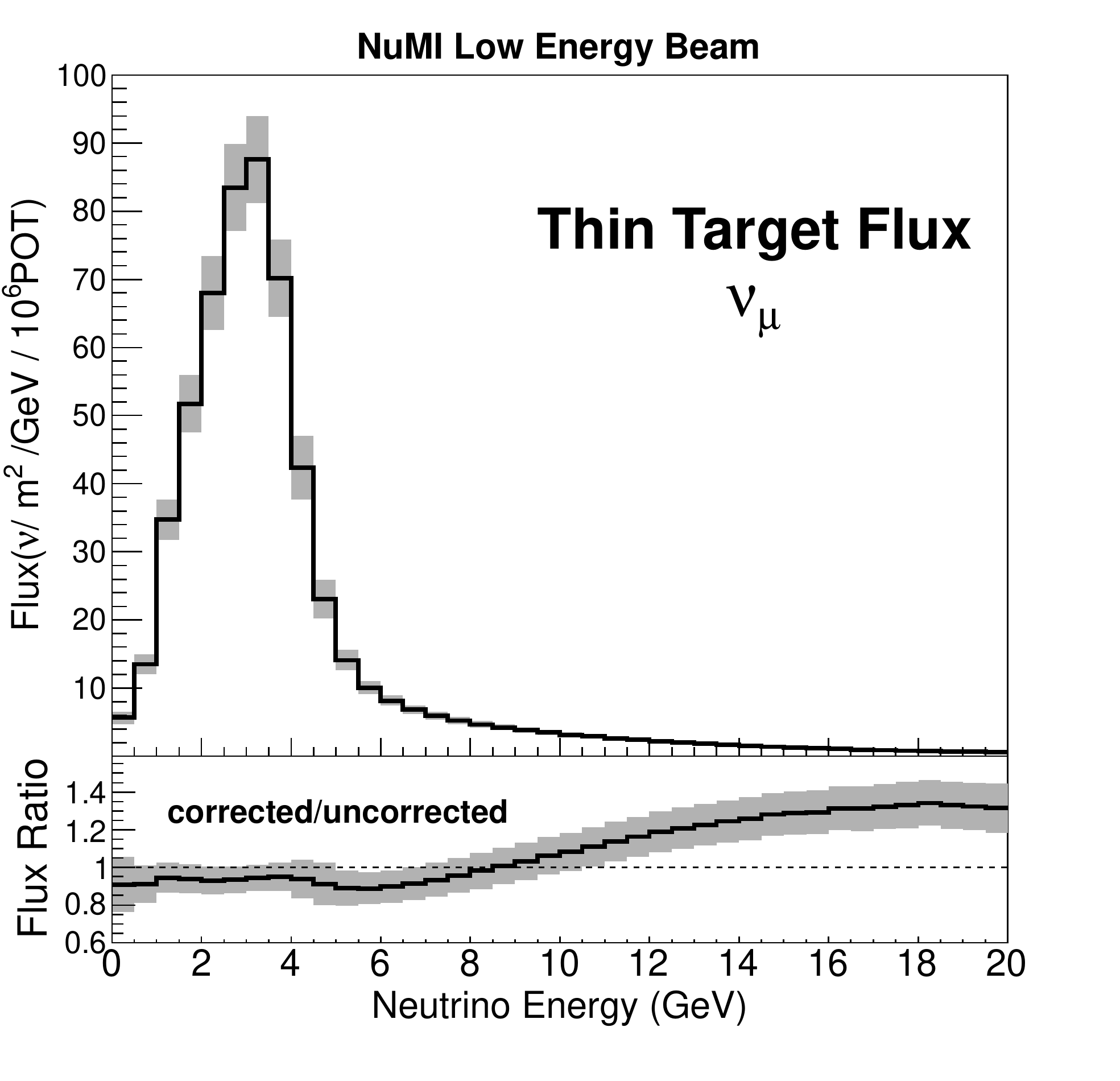}
\caption{The predicted thin target ~\numu flux at the MINERvA detector for the low energy, \numu focused, beam configuration. The ratio plot shows the effect of correcting the flux simulation using thin target hadron production and attenuation data as described in the text. The error band includes uncertainties due to hadron interactions, beam geometry and beam focusing.}
\label{fig:le_thin_flux}
\end{figure}

The  thin target flux prediction for the NuMI LE \numu beam is shown in Fig.~\ref{fig:le_thin_flux}. The prediction uses thin target data combined with the {\em ab initio} uncertainty estimates for process that lack a data constraint. The ratio between the corrected and uncorrected flux predictions is also shown.  Incorporating the corrections described here reduces the predicted flux near the focusing peak by 8\% while in the high-energy region it increases the prediction by as much as 30\%. The uncertainties as a function of neutrino energy, which were shown separately for hadron production and beam focusing in Fig.~\ref{fig:thin_unc}  and Fig.~\ref{fig:foc_unc}, are combined in this procedure and shown in the error bands in Fig.~\ref{fig:le_thin_flux}.  The \numu flux is $\unit[287\pm 22]{\nu_\mu /m^{2}/10^{6}\mathrm{POT}}$ when integrated over the $0< E_\nu <\gev{20}$ range.

%The \numu flux is $\unit[287\pm 22]{\nu_\mu m^{-2} 10^{-6} \mathrm{POT}^{-1}}$ when integrated over the $0< E_\nu <\gev{20}$ range.

\begin{figure}
\centering
\includegraphics[width=\columnwidth]{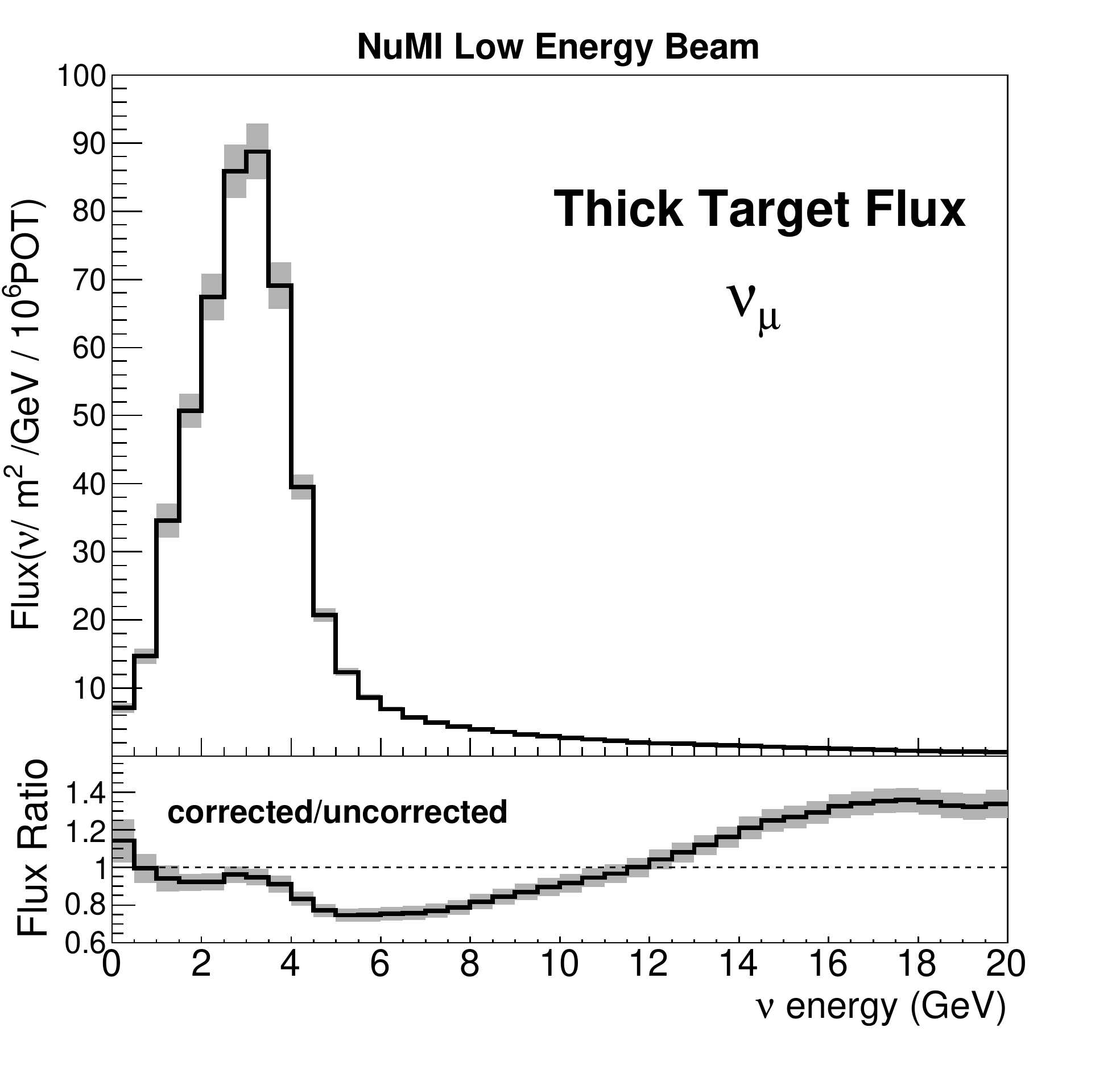}
\caption{The predicted thick target ~ \numu flux at the MINERvA detector for the low energy, \numu focused, beam configuration. The ratio plot shows the effect of correcting the flux simulation using thick and thin target hadron production and attenuation data as described in the text. The error band includes uncertainties due to hadron interactions, beam geometry and beam focusing.}
\label{fig:le_thick_flux}
\end{figure}

The thick target flux prediction for the NuMI LE \numu beam is shown in Fig.~\ref{fig:le_thick_flux}. The data used in the prediction are predominantly the $\pi$ and $K$ yields measured by MIPP. The prediction also uses some thin target data to fill in gaps, as well as the {\em ab initio} uncertainty estimates on processes that lack a data constraint. Table~\ref{tab:thickcoverage} shows the fraction of thin and thick target data used in the thick target ~prediction.

\input{table2.tex}

%\subsubsection*{A comparison of thick and thin fluxes}

\begin{figure}
\centering
\ifnum\DoPrePrint=0
\includegraphics[width=\columnwidth]{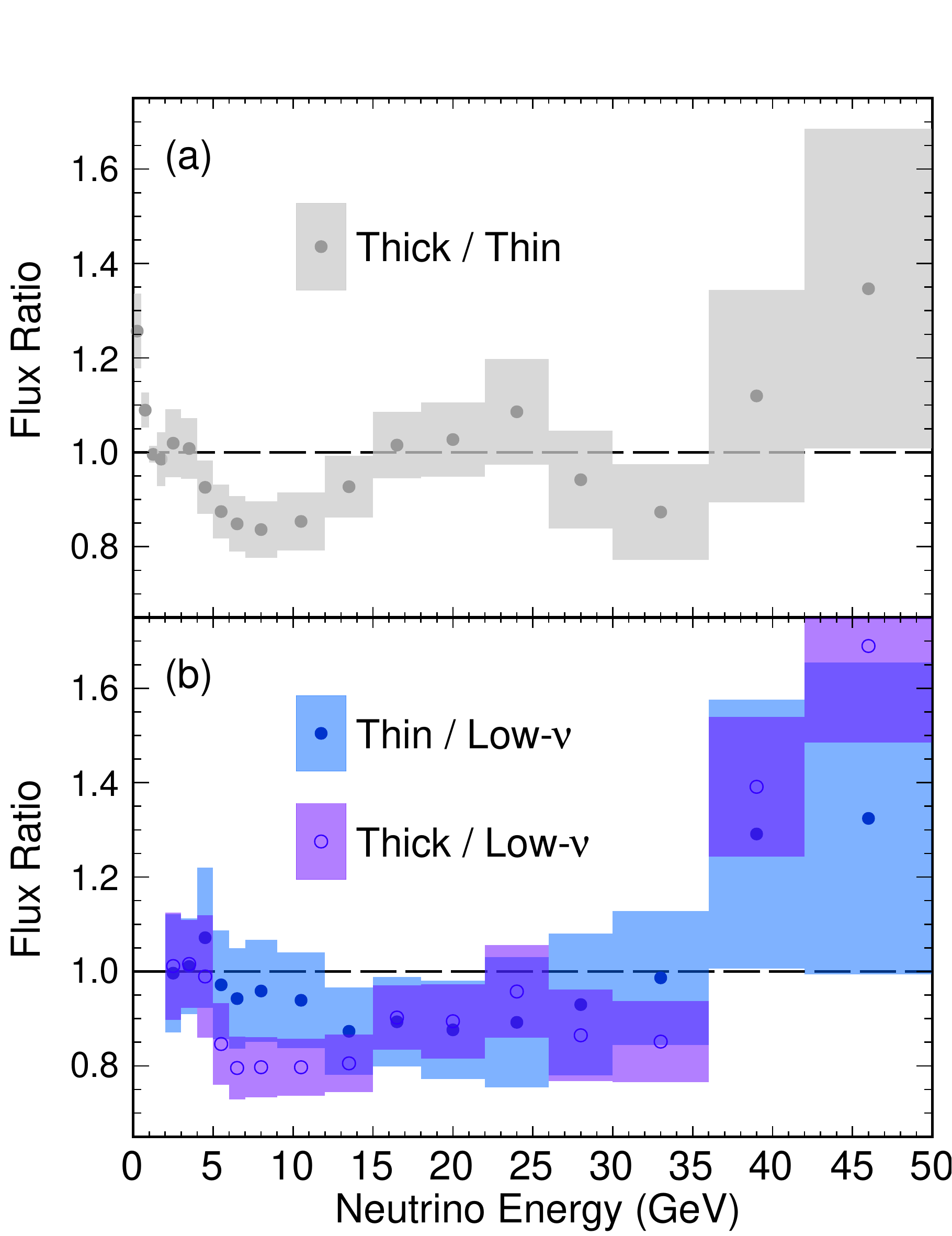}
\else
\centering
\includegraphics[width=.85\columnwidth]{thick_thin_lownu_ratios}
\fi
\caption{Ratios of flux predictions. (a) The flux predicted using data from thick target experiments divided by the flux prediction that uses only thin target data. (b) The thin and thick target flux predictions divided by the {\it in situ} flux measured using the \lownu technique. The error bands on each curve account for uncertainties in the numerator and denominator, including the effect of significant correlations between the thick and thin target predictions.}
\label{fig:thick_thin_lownu}
\end{figure}

Figure~\ref{fig:thick_thin_lownu}(a) shows the ratio between the thin and thick target flux predictions. The error band and covariance matrix were constructed using the multi-universe technique and account for correlations between systematic uncertainties that are common to the two predictions. There is a clear discrepancy between the two, especially in the region $4\lesssim E_\nu \lesssim \gev{15}$. This is due to a large suppression, relative to the simulation, of \pip yields in the thick target data in the range $10\lesssim p_{z} \lesssim \gev{40}$. We quantified the level of agreement by computing the $\chi^2$ between the two predictions:
\begin{equation}
    \chi^2_{tt} = \sum_{i,i\leq j}^{N} ( \phi^{thick}_i -\phi^{thin}_i ) (\phi^{thick}_j - \phi^{thin}_j) [\mathbf{V^{-1}_{tt}}]_{ij}
\end{equation}
where $\phi^{thick,thin}$ refers to the flux predictions in the bins $i,j$ and $\mathbf{V_{tt}}$ is the bin-to-bin covariance matrix.  For $0<E_\nu<\gev{50}$ we find $\chi^2_{tt}=144.7$ for 19 degrees of freedom ($p=10^{-21}$). MINERvA cross section analyses tend to include events in the energy range of $2< E_\nu<22$~GeV, and we find $p=10^{-12}$ for this range.

%% file: table2.tex
\begin{table*}[]
\centering
\begin{tabular}{|c|d|d|d|d|d|}
\hline
$E_\nu$ (GeV) & 3-4 & 6-7 & 9-10 & 14-15 & 19-20 \\
\hline
thick (\%) & 87 & 76 & 70 & 69 & 75 \\
thin \& {\em ab initio} (\%) & 13 & 24 & 30 & 31 & 25 \\
\hline
\end{tabular}
\caption{The impact of thin and thick target datasets on the prediction of the ``thick target'' \numu flux at the MINERvA detector in the LE beam configuration. The rows show the fraction of interactions which lead to a \numu that are covered by the thick and thin target datasets, including the {\em ab initio} uncertainty estimates made for some processes.}
\label{tab:thickcoverage}
\end{table*}

%% file: insitu.tex
%\section{Comparisons with {\it in situ} data.}

\minerva has two {\it in situ} flux constraints that can in principle help discriminate between the thin and thick target predictions. First, the rate of $\nu e^- \to \nu e^-$ was measured with a precision of 11.5\% and can be used to constrain the flux since the process has a well known cross section~\cite{nuepaper}. The measurement is sensitive to the integrated flux but only weakly sensitive to the $E_\nu$ dependence since only the electron energy can be measured in the detector and the outgoing neutrino carries away significant energy. The measured rate is in good agreement with both the thin- and thick target predictions, mostly because the LE flux is peaked in the range $2<E_\nu < \gev{5}$ where the two predictions differ by less than the statistical precision of the $\nu e^-$ scattering measurement.  

The second {\it in situ} constraint uses a sample of $\numu A \to \mu^- X$ events in which the energy of the recoil system ($\nu$) is much lower than the neutrino energy~\cite{joshthesis}. The cross section for this ``\lownu'' ~process has a weak dependence on the neutrino energy that is understood at the few per cent level~\cite{ariepaper}.  Therefore, this event sample can be used to predict the energy dependence of the flux. The overall level of flux is then determined by computing the \numu charged current scattering cross section using an inclusive sample of $\numu A \to \mu^- X$ events, and requiring that it matches the NOMAD measurement at $E_\nu = \gev{10}$ which has an uncertainty of 3.6\%~\cite{nomad}.

Figure~\ref{fig:thick_thin_lownu}(b) shows a comparison of the thick- and thin target flux predictions divided by the \lownu flux measurement. We have again quantified the level of agreement using a $\chi^{2}$ test. The results are shown in Table~\ref{tab:chi2}.  The thin target and \lownu fluxes agree well but the agreement for the thick target flux is poor (for the 2-\gev{50} range) to marginal (2-\gev{22}).  We also tabulated the likelihood ratio $r=L_{thick}/L_{thin} = \exp\left[-\frac{1}{2} \left(\chi^2_{thick} - \chi^{2}_{thin}\right)\right]$.

\begin{table}[h]
\centering
\begin{tabular}{|l|c|c|c|c|c|c|}
\hline
 $E_\nu$ Range:& \multicolumn{3}{c|}{2-50 GeV} & \multicolumn{3}{c|}{2-22 GeV} \\
 \hline
flux comparison & $\chi^2$ & NDF & $p-value$   & $\chi^2$ & NDF & $p-value$ \\
\hline
thin-\lownu & 7.3 & 15 & 0.95 & 4.8 & 10 & 0.91 \\
\hline
thick-\lownu & 61.3 & 15 & $1.5\times10^{-7}$ & 18.6 & 10 & $4.6\times 10^{-2}$ \\
\hline
$r=\frac{L_{thick}}{L_{thin}}$ & \multicolumn{3}{c|}{$2\times 10^{-12}$} &\multicolumn{3}{c|}{$1\times 10^{-3}$} \\
\hline
\end{tabular}
\caption{Results from a $\chi^2$ comparison of the thick- and thin target constrained fluxes with the \lownu flux.}
\label{tab:chi2}
\end{table}

In principle yet a third flux prediction could be found by combining the thin- and thick target predictions. However, because the two predictions disagree, combining them would require increasing the uncertainties in the measurements appropriately.  Because the likelihood ratio $r$ strongly disfavors the thick target flux, we chose not to combine the two predictions. The thin and thick target predictions for the \numubar flux in the \numubar enhanced beam configuration are in good agreement, and agree with the \lownu constraint. We have chosen not to combine the two \numubar flux predictions at this time. 

%assumptions must be made on we do not think that a combination would produce a more accurate result. Moreover, the precision would decrease since the discrepancy between the two predictions %would have to be incorporated as an additional systematic uncertainty. Instead, we note that the likelihood ratio $r$ strongly disfavors the thick target flux.

%However, since the $\nu e$ scattering measurement of the flux is an independent check, we can use that measurement to provide still more precision, as described in Ref.~\cite{nuepaper}.  
%with a significance of $2\times 10^{-12}$ for the 2-\gev{50} range and $1\times 10^{-3}$ for the 2-\gev{20} range. 
%We therefore have decided to use the thin target prediction constrained by our electron scattering measurement for cross section analyses~\cite{jeremy}. 
%while continuing to study the thick target prediction. 

Because the $\nu e^-$ scattering measurement of the integrated flux~\cite{nuepaper} is independent of the measurement here, it can be used to further improve the precision of cross section measurements in MINERvA.  We will use the thin target prediction presented here, as constrained by the $\nu e^-$ measurement, for future cross section analyses. The flux for all neutrino species in the low energy \numu-enhanced, and \numubar-enhanced beams is provided as supplemental material.

%% file: conclusions.tex
\section{Conclusions} 

This paper presents the first {\em a priori} prediction of the NuMI low energy flux. MINERvA's published cross-section results have used early forms of this prediction which are now superseded. The flux reported here is the most precise {\em a priori} prediction available given the current state of hadron production measurements and the constraints coming from the {\em in situ} flux measurements. A previous ``neutrino independent'' constraint that used the NuMI muon monitoring system had uncertainties of 10-25\% over the 0-\gev{25} range~\cite{laurathesis}.  The uncertainty on the thin target flux prediction is 7.8\% when the flux is integrated from 0 to \gev{20}. Hadron production uncertainties dominate the flux uncertainty, except in the region around \gev{5} where beam focusing uncertainties are important. The uncertainty on the thick target flux integrated over the same range is 5.4\%, demonstrating the value of dedicated hadron production measurements using actual or replica targets. The discussion in this article has focused on the NuMI beam but the technique of constraining a flux prediction with hadron production and interaction measurements can be applied to other similar beams, in particular the Long Baseline Neutrino Facility that will provide the beam for the DUNE experiment. 

%~\cite{*[{A previous a priori prediction using the NuMI muon monitor system and an in situ measurement using the low-nu method are documented in }] [{ respectively.}] laurathesis}.

%MINERvA has already collected well over three times the integrated protons on target in the ME beam compared to the earlier run described here, and 
%%and given the improvements in focusing and increase in cross section 
%the total statistics for the {\em in situ} events are between a factor of 6 and 8 above what were accumulated in the LE beam.  These statistics will enable a substantial increase in the precision of the \lownu measurement and also of the $\nu e \to \nu e$ flux measurement, transforming the latter from a cross-check at comparable precision into a strong constraint. 
%%\nova is also pursuing a $\nu e \to \nu e$ measurement with we expect will complement the \minerva result. 

%The USNA61 collaboration has proposed hadron production measurements on thin carbon and aluminum targets as well as on a replica of the medium energy NuMI target~\cite{usna61}. Obtaining high precision thick target data is extremely important and may be the only way to reduce {\it a priori} flux uncertainties below $\sim 5\%$. 

%Ref.~\cite{mumons} describes efforts underway to measure the tertiary muons in NuMI which
%could provide an {\em in situ} flux constraint that does not depend on neutrino interactions.  

%% file: acknowledgments.tex
\begin{acknowledgments}

%{\LARGE I don't know if these are the latest:  they are the same as the nu_e ccqe}\\

This work was supported by the Fermi National Accelerator Laboratory
%, which is operated by the Fermi Research Alliance, LLC, 
under US Department of Energy contract
No. DE-AC02-07CH11359 which included the \minerva construction project.
Construction support also
was granted by the United States National Science Foundation under
Award PHY-0619727 and by the University of Rochester. Support for
participating scientists was provided by NSF and DOE (USA) by CAPES
and CNPq (Brazil), by CoNaCyT (Mexico), by CONICYT (Chile), by
CONCYTEC, DGI-PUCP and IDI/IGI-UNI (Peru), by Latin American Center for
Physics (CLAF).
% and by RAS and the Russian Ministry of Education and Science (Russia).  
We thank the MINOS Collaboration for use of its
near detector data. Finally, we thank 
Fermilab for support of the beamline and the detector, and in particular the Scientific Computing Division and the Particle Physics Division for support of data processing.  \end{acknowledgments}

%% file: main.bbl
\begin{thebibliography}{45}
\bibitem[{nuc()}]{nucleonA}
\bibinfo{note}{Note: We categorize $pC \to p X$ interactions at $x_{F}>0.95$ in
  the nucleon-A category for technical reasons. They comprise 20\% of the
  category.}
\bibitem[{na4()}]{na49_systematic}
\bibinfo{note}{NA49 doesn't provide a covariance matrix with their results.
  But, approximately half of the experiment's systematic error is due to
  normalization, indicating at least a 50\% correlation between each point.
  Some of the other uncertainties may also have significant bin to bin
  correlations. We have assumed 100\% because that is conservative,
  operationally easy, and because the NA49 systematic uncertainty is rather
  small.}
\bibitem[{mip()}]{mipp_systematic}
\bibinfo{note}{MIPP doesn't provide a covariance matrix with their results. A
  75\% correlation is our best guess after speaking with the MIPP authors and
  erring on the conservative side. We are not very sensitive to this
  assumption. For example, the uncertainty on the flux at 3~GeV is about 3.9\%
  if we assume a 75\% correlation, about 4.6\% if we assume a 99\% correlation,
  and about 3.2\% if we assume a 50\% correlation.}
\end{thebibliography}
